# Magnetization Relaxation via Quantum and Classical Vortex Motion in a Bose Glass Superconductor

Leo Radzihovsky

*The James Franck Institute and Physics Department, University of Chicago, Chicago, IL 60637*

(April 8, 1995)

I show that in Bose Glass superconductor with high $j_c$ and at low $T$ the magnetization relaxation (S), dominated by quantum tunneling, is $\propto \sqrt{j_c}$, which crosses over to the conventional classical rate $\propto T/j_c$ at higher $T$ and lower $j_c$, with the crossover $T^* \sim j_c^{3/2}$. I argue that due to interactions between flux lines there exist three relaxation regimes, depending on whether $B < B_\phi$, $B \approx B_\phi$, $B > B_\phi$, corresponding to Strongly-pinned Bose Glass (SBG) with large $j_{c2}$, Mott Insulator (MI) with vanishing S, and Weakly-pinned Bose Glass (WBG) characterized by small $j_{c1}$. I discuss the effects of interactions on $j_c$ and focus attention on the recent experiment which is consistently described by the theory.

74.60.Ge, 74.60.Jg, 74.20.Hi

In high $T_c$ superconductors flux line vortices play a major role in determining dissipation effects, and as a result of strong fluctuations a fairly passive Abrikosov lattice is replaced by a variety of new phases. At high temperatures and fields, thermal fluctuations melt the Abrikosov lattice over a large portion of the B-T phase diagram resulting in a strongly dissipating vortex liquid [1]. On the other hand, quenched disorder, at low T and B, leads to a variety of proposed glassy phases with Vortex Glass [2] corresponding to point disorder such as oxygen vacancies and interstitials, and Bose Gass [3] resulting from the introduction of columnar defects into the superconductor [4].

In the presence of both point and correlated disorder there is a variety of regimes with very different mechanism for flux line motion, depending on the relative strength of disorder, temperature, and current, many of which have been lucidly discussed in Refs. [2,3]. In this Letter I focus on a narrower range of behavior, relevant to the specific magnetization-relaxation experiment that will be discussed below [5]. I study vortex motion at high currents $j \approx j_c$ and very low temperatures, deep in the Bose Glass regime, by adapting the quantum creep theory [6] to dissipation in the presence of columnar defects. I show that quantum relaxation can be distinguished from the classical regime not only by the absence of temperature dependence but also by the counter-intuitive increase of the magnetization relaxation rate with $j_c$, which I find to be $\propto \sqrt{j_c}$ in the quantum dissipation-dominated regime.

Studies of magnetization relaxation have provided us with important clues into the mechanism of flux line motion in the presence of currents and disorder. There appears a natural separation of two time scales. An exponentially fast relaxation of B to the Bean critical state profile [7] takes place as long as local currents exceed the critical current $j_c$, determined by the strength of pinning of vortices, followed by a very slow, typically $\log t$ relaxation, the regime that I focus on in this work.

Our understanding of the motion of vortices goes back to the Anderson-Kim (AK) model [8]. The flux lines are treated as independent particles thermally activated over finite impurity barriers $U$, moving down a washboard potential representing disorder with the average slope proportional to the external current $j$. Because the barriers are finite and flux lines therefore can be thermally activated over the local barrier, this theory predicts a finite linear flux flow resistivity for $j \gg j_c$ and for $j \ll j_c$ (although strongly suppressed by $\exp(-U/T)$ factor). Near $j_c$ the I-V characteristic is highly nonlinear with $V(j) \sim \exp(j/j_c - 1)^\zeta$. Recently, however, it has been appreciated that for $j < j_c$ in glass phases the vortex motion is via vortex loop nucleation over barriers. This leads to effective barriers $U(j) \sim (1/j)^\mu$ that diverge as $j \to 0$, where $\mu$ is determined by the details of the glass phase and by the range of $j$. [2,3,9] Although these theories disagree with AK theory at low $j$, predicting vanishing linear resistivity, all the classical models of vortex motion, for $j \leq j_c$, lead to the following form of magnetization relaxation out of the Bean state,

$$M(t) \approx M_o \left[ 1 + \frac{T\mu}{U} \log(t/t_o) \right]^{-1/\mu} \quad (1)$$

which leads to classical magnetization relaxation rate

$$S_c = -\frac{1}{M_o}\frac{dM}{d(\log t)} \approx \frac{T}{U} \quad (2)$$

An experimentally verifiable signature of this classical prediction is that $S_c$ vanishes linearly with $T$ and is inversely proportional to $j_c = Uc/\phi_o b$, where $\phi_o = hc/2e$, $b$ is the effective diameter of the columnar defect and both $U$ and $b$ are likely to be renormalized by the thermal fluctuations and the interactions.

In numerous recent experiments at very low $T$, however, the creep rate appears to extrapolate to a finite value at $T = 0$ [10] which has been taken as evidence of quantum tunneling of vortices out of pinning sites [6,11,12]. Using instanton methods it can be shown [13] that the thermal depinning rate $\exp(-U/T)$ is replaced by the tunneling rate $\exp(-S_E/\hbar)$ where $S_E$ is the Euclidean action evaluated at the stationary field configuration. In analogy to Eq.2, the quantum magnetization relaxation is given by [6]

$$S_q \approx \frac{\hbar}{S_E} \quad (3)$$

an appealing result since $\hbar$ analogously to $T$ controls quantum fluctuations and $S_E$ plays the role of a quantum barrier in $3+1$-space-time dimensional field theory.

A transparent way to understand the form of $S_E$ is to look at the equation of motion for the flux line

$$M\frac{d^2\vec{r}(z,t)}{dt^2} + \alpha(\frac{d\vec{r}(z,t)}{dt} - \vec{v}_s) \times \hat{z} + \eta\frac{d\vec{r}(z,t)}{dt} = -\frac{\delta F[\vec{r}]}{\delta\vec{r}(z,t)} \quad (4)$$

where $\vec{r}(z,t)$ is the displacement of the flux line transverse to $z$-axis, $M$ is the effective vortex mass, $\eta$ and $\alpha$ are the viscous and Hall drag coefficients, $\vec{v}_s$ is the local superfluid velocity, and $F$ is the vortex free energy. In the above equation, the first two terms are the nondissipative inertial and Hall forces, and the third term is the Bardeen-Stephen dissipative friction force, with $\eta \approx \phi_o H_{c2}/\rho_n c^2$. It is not difficult to include all three terms in the analysis, however in high $T_c$ materials for $j \approx j_c$ and at long times I expect that the viscous drag term will dominate and I therefore focus on this term [14]. The effects of the nondissipative inertial and Hall terms have been studied in Refs. [6,11] for point disorder and by me in the case of columnar defects. The inertial term leads to a $T$ and $j_c$-independent contribution to $S_q$, while the Hall term (by dimensional analysis) has the same scaling as the viscous contribution. The functional integration over $\vec{r}$ (analytically continued to imaginary time and transformed to frequency domain) of the resulting simplified equation leads to the Euclidean action

$$S_E = \int dwdz \left[\frac{\eta}{2}|w||\vec{r}(w,z)|^2 + \frac{\epsilon_l}{2}\left(\frac{\partial\vec{r}}{\partial z}\right)^2 + U(\vec{r}) - \vec{f}_L\cdot\vec{r}\right] \quad (5)$$

where the second term is the elastic energy, $U(\vec{r})$ is the pinning energy, and the last term is the contribution due to the Lorentz force, $\vec{f}_L = \phi_o\hat{z}\times\vec{j}/c$, acting on the flux line. To be specific I consider the regime in which the coherence length $\xi \approx 15\text{Å}$ is significantly smaller than the diameter of the columnar defect $b \approx 70\text{Å}$. In this case, $U(\vec{r}) = -U \approx -\epsilon_o \equiv -(\phi_o/4\pi\lambda)^2$, for $r < b$ and vanishes for a flux line outside of the columnar defect. For $B \gg \phi_o/\lambda^2$, when the average spacing between flux lines $a = (\phi_o/B)^{1/2} \gg \lambda$, the tilt modulus is well approximated by $\epsilon_l \approx \gamma\epsilon_o$, with $\gamma = (M_\perp/M_z)\log(a/\xi)$ describing material anisotropy. The viscous drag term ($\eta$) in Eq.5, is identical to the contribution to the effective action introduced previously by Caldeira and Leggett [15] to account for dissipation in a quantum system.

To compute the quantum relaxation rate $S_q$, I first determine the space-time field configuration of the tunneling vortex loop, characterized by the $l_\perp$, $l_z$ space length scales and time scale $t_o$, by balancing terms in the action. Balancing the pinning energy ($U = b\phi_o j_c/c$) against the Lorentz energy, $Ul_z \approx j\phi_o l_z l_\perp/c$, leads to $l_\perp \approx b(j_c/j)$, which in a magnetization relaxation experiment with $j \approx j_c$ reduces to $l_\perp \approx b$. The relaxation is therefore dominated by half-loops and only involves individual columnar pins in each nucleation process [3]. Competition between the elastic and Lorentz energies, $\epsilon_1 b^2/l_z \approx j_c\phi_o l_z b/c$, gives $l_z \approx b(j_1/j_c)^{1/2}$, where $j_1 = \epsilon_1 c/\phi_o b$ is proportional to the depairing current $j_0$. Finally, balancing these free energy contributions against the dissipation term, $\eta b^2/t_o \approx \epsilon_1 b^2/l_z^2$ gives $t_o \approx (\eta/\epsilon_1)l_z^2$. Substituting this instanton configuration inside $S_E \approx \epsilon_1 t_o b^2/l_z$ and using Eq.3 I obtain,

$$S_q = \left(\frac{\xi}{b}\right)^2 \left(\frac{\rho_n}{b}\right)\left(\frac{e^2}{\hbar}\right)\sqrt{j_c/j_1} \quad (6)$$

The above result obtained for individual columnar defect pinning is surprisingly similar to the case of collective quantum creep for point disorder, previously obtained by Blatter, et al. [6]. This Bose Glass result is suppressed by the factor of $(b/\xi)^3$ relative to the point disorder result, and the depairing current $j_o$ is replaced by $j_1 \approx \gamma(\xi/b)j_o$. This counter-intuitive dependence of $S_q$, in which the relaxation rate increases with the pinning strength (larger $j_c$) is a clear and intriguing sign of quantum vortex motion.

The total magnetization relaxation rate $S$ is a combination of quantum and classical rates, which have very different $j_c$ and temperature dependences. Although a detailed crossover function is difficult to compute, much of the interesting information can be obtained from the asymptotic behavior of $S$

$$S = a_{qn} + a_{qd}\sqrt{j_c}, \text{ for } T < T^*$$
$$= a_c\frac{T}{j_c}, \text{ for } T > T^*, \quad (7)$$

where $a_{qd}$ is defined by Eq.6, $a_c \approx c/b\phi_o$ (see Eq.2), and $a_{qn}$ is $j_c$-independent contribution due to quantum nondissipative inertial term that I expect to be small [6]. The crossover as a function of temperature from constant quantum to linear classical relaxation previously has been observed in several experiments. Balancing the quantum and classical relaxation rates in Eq.7 predicts the crossover temperature $T^* \sim j_c^{3/2}$. Furthermore, if a consistent study of $S$ as a function of $j_c$ can be made, I predict that quantum $\sqrt{j_c}$ dependence dominant at high critical currents will crossover to classical $1/j_c$ behavior in samples with smaller $j_c$, and this will occur at $j_c^* \sim T^{2/3}$. In addition, a crossover as a function of $j_c$ from a dissipative to a nondissipative regime can take place at low T and $j_c$ (see Eq.7).

The dissipative quantum relaxation regime predicted by Eq.7 and a hint of crossover to classical relaxation as a function of $j_c$ might have already been observed in recent experiments on YBCO single crystals at milliKelvin temperatures. The samples were irradiated with heavy ions to create a forest of columnar defects with density characterized by $B_\phi = \phi_o/d^2$ (where $d$ is the average distance between the defects), and the magnetization relaxation has been measured [5]. In Figure 1 I displayed a smooth interpolation between the quantum and classical limits of Eq.7 and fit it to the $S(j_c)$ data that I extracted from the experiment. At higher $j_c$ the data clearly shows a counter-intuitive increase in $S$ as the pinning strength ($j_c$) is increased, which I could not explain by any classical mechanism, in agreement with the dissipative quantum tunneling predicted by Eq.7. The flattening out at low $j_c$ is also consistent with the crossover to classical relaxation, as discussed above.

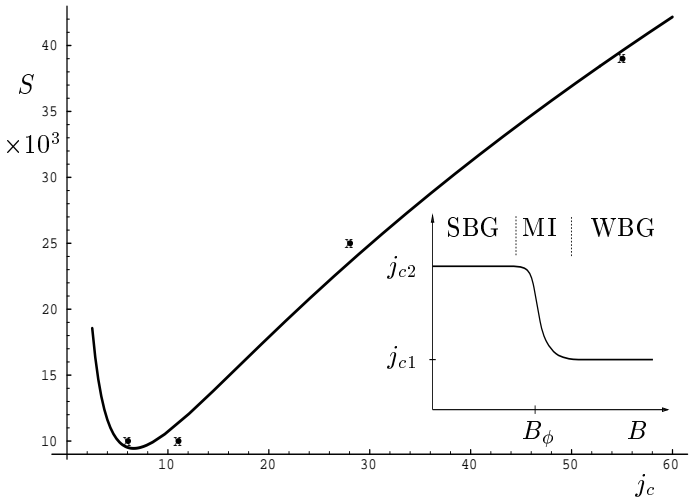

FIG. 1. Relaxation data at $T = 0.1$ taken from Ref.5, plotted as a function of $j_c$. The high (low) $j_c$ pair of points corresponds to data in SBG, $B \ll B_\phi$ (WBG, $B \gg B_\phi$) regimes for $B_\phi = 1$ and 2 Tesla samples, respectively. The solid curve is a smooth interpolation between the quantum and classical limits of Eq.7, fit to the experimental data. The inset shows the $j_c$ step-like behavior as a function of B predicted by the existence of SBG and WBG regimes, consistent with the experiment.

Although this quantum relaxation calculation Eq.7 appears to explain the qualitative feature of the intriguing increase of $S$ with $j_c$ observed in the experiment of Ref. [5], an important puzzle remains. My calculation is only strictly correct for the regime where the pinning is dominated by individual defects. While this regime is appropriate to the range of fields explored in the experiment of Ref. [5], it is difficult to simultaneously reconcile the observed increase of $j_c$ (pinning strength) with the increased density of pins ($B_\phi$) in this single defect pinning regime. It is clear that in Eq.7 the length scale $b$ and pinning energy $U = b\phi_o j_c/c$ should be interpreted as the effective pinning length and pinning energy, renormalized by flux line interaction and temperature. However, it is not at all clear how these renormalized quantities are affected by the density of columnar defects, i.e. their dependence on $B_\phi$. In the absence of a detailed theory of this dependence, I take the increase with $B_\phi$ of the renormalized pinning energy $U$ ($j_c$) appearing in Eq.7 as an experimental fact (that requires further investigation). Assuming that the change in density of defects has weaker affect on $b$ than the observed increase of $j_c$, Eq.7 provides a reasonable explanation for the counter-intuitive experimental observation of the increase of $S$ with $j_c$.

I now turn to the effects of interactions between flux lines on the above picture. In examining the role of interactions on pinning in the Bose Glass phase I have found two quite distinct pinning mechanisms, for $B < B_\phi$ and $B > B_\phi$ to which I refer as Strongly-pinned Bose Glass (SBG) and Weakly-pinned Bose Glass (WBG), respectively. For $B < B_\phi$ the columnar defects outnumber the flux lines, and therefore all the vortices are strongly pinned by individual defects. The pinning energy in this case is of order $U \approx \epsilon_o \log(b/\xi)$ and $j_{c2} \approx Uc/\phi_o b$. As $B$ is increased past $B_\phi$, a competition between pinning energy and magnetic repulsion ensues, resulting in lowering of the effective pinning potential. The additional flux lines can attempt to double-occupy the already occupied columnar defects, thereby gaining condensation energy $U$, but raising the interaction energy by $V_{int} \approx 2\epsilon_o \log(a/b)$. In these experiments $a \approx 400$Å, $b \approx 70$Å, the magnetic repulsion dominates and the additional flux lines go into the interstitials in between the vortices localized on the columnar defects. Hence for $B > B_\phi$ the weakly pinned interstitials flux lines are localized by magnetic repulsion from the vortices strongly pinned by the columnar defects. Although the pinning potential seen by the interstitial vortices is not isotropic in the ab-plane, because the potential is $z$-independent, the Bose Glass phase is expected to persist for $B > B_\phi$, as was first argued in Ref. [3]. I expect this WBG phase, however, to have a significantly reduced critical current, $j_{c1}$ relative to that of the SBG phase.

The step-like behavior of $j_c$, shown qualitatively in the inset of Fig.1, has necessarily experimentally verifiable consequences that the spatial linear field profile of the Bean state will be replaced by a two slope configuration corresponding to $j_{c1}$ and $j_{c2}$, with the change in the slope occurring at $x(B_\phi)$. The measurements of the hysteresis loop already find $j_c(B < B_\phi) = j_{c2} >> j_c(B > B_\phi) = j_{c1}$ in the strong support of the existence of distinct SBG and WBG pinning regimes. [5]

I further observe that in SBG regime both $B_\phi = 1$ and 2 Tesla samples studied in Ref. [5] exhibit a temperature independent relaxation rate up to 4 Kelvin, giving additional evidence for quantum creep mechanism. On the other hand in the WBG regime, with significantly lower $j_c$, the relaxation rate is approximately linear with temperature down to $T^* \approx 0.4K$, manifesting a classical relaxation mechanism. This is in qualitative agreement with Eq.7 which gives a quantum, $T$-independent rate for high $j_c$ and a classical, linear in $T$ rate for low $j_c$.

Another striking observation is the strong suppression of $S$ over a narrow range of $B$ around $B \approx B_\phi$. The location of the dips in $S$ suggests that the experiment is the first observation of the Mott Insulator phase first discussed in Ref. [3]. In this regime the number of flux lines matches the number of columnar defects, and due to magnetic repulsion between vortices the hopping rate is expected to be significantly reduced. Although the qualitative picture is clear, a more thorough understanding of the interactions is needed to describe $S$ near $B_\phi$ in detail.

I now examine the behavior of $j_c$ as a function of the columnar defect density $B_\phi$. I note that $j_c$ is observed to approximately double as $B_\phi$ is increased from 1 to 2 Tesla. This is expected since higher density of columnar defects increases the average pinning strength by the ratio of attractive pinning area to total area $(b/d)^2 \propto B_\phi$. Flux line interactions lead to another source of $B_\phi$ dependence in $j_c$. Since in the absence of defects the flux lines would like to form an equally spaced lattice, in order to take advantage of the columnar defect, a flux line must displace an average distance $d/2$ from its otherwise preferred position. Hence the effective energy gain from localizing on the defect is reduced by the increase in the repulsion with $(\lambda/a)^2$ other lines to $U_{eff} \approx U - \delta\epsilon_o(d/2a)^2(\lambda/a)^2 \approx \epsilon_o(1 - \delta B^2/4B_\phi B_{c1})$ (for $B << B_\phi$, $a^* < \lambda$). This dependence predicts an additional slower increase in $j_c$ with $B_\phi$ and leads to a critical field $B^* \approx 2\sqrt{B_\phi B_{c1}/\delta}$ above which the effective pinning strength vanishes. In the case $a^* > \lambda$, $B^* \approx B_\phi$ [3]. Turning the argument around, I therefore expect a critical $B_\phi^c(B)$ density of defects for a realization of a true Bose Glass phase.

In summary, I have discussed a theory of magnetization relaxation at ultra low temperatures in a Bose Glass superconductor, predicted a quantum to classical crossover as a function of $j_c$ as a new signature of quantum relaxation in addition to the usual $T$-dependence signature, and have argued for the existence of SBG and WBG distinct pinning regimes. I have examined recent experiments, and found that they can be consistently explained in the context of the developed picture.

Note added: After this paper was submitted for publication, one of the referees alerted me of the existence of the paper by V. Vinokur, Physica A Vol. 200, p. 384 (1993), which also explores quantum vortex tunneling from a columnar defect. My findings are in disagreement with Vinokur's, who finds that the relaxation rate (in the regime considered here) is independent of $j_c$, inconsistent with experiments of Ref. [5].

This research was supported by the NSF (DMR 94 − 16926), through the Science and Technology Center for Superconductivity. I thank K. Beauchamp and T. Rosenbaum for helpful discussions and for providing their data before publication. I also thank an anonymous referee for constructive criticism.